\documentclass[11pt,twoside]{article}
\usepackage{newpasp}

\usepackage{epsf}
\usepackage{graphicx}

\index{Young, L. M.}
\index{van Zee, L.}
\index{Dohm-Palmer, R. C.}
\index{Lo, K. Y.}

\begin{document}

\title{Star Formation and the ISM in Dwarf Galaxies}
\author{L. M. Young}
\affil{Physics Dept., New Mexico Tech, Socorro NM 87801}
\author{L. van Zee}
\affil{Herzberg Institute of Astrophysics, Victoria, BC, Canada}
\author{R. C. Dohm-Palmer}
\affil{Astronomy Dept., University of Michigan, Ann Arbor, MI 48109}
\author{K. Y. Lo}
\affil{Institute of Astronomy and Astrophysics, Academia Sinica, Taipei, Taiwan ROC}


\begin{abstract}
High spatial and spectral resolution observations of the atomic 
interstellar medium in nearby dwarf galaxies reveal
evidence for warm and cold neutral gas, just like the phases in our
own Galaxy.
The cold or quiescent phase ($\approx$ 20\% of the HI in the galaxies studied,
except for LGS~3) seems to
be associated with star formation activity--- it may mark the regions
where the conditions are right for star formation.
These results help to explain the patterns of star formation activity which
are seen in color-magnitude data for the dwarf irregulars.
\end{abstract}





\section{Motivation}
Color-magnitude diagrams of the stars within the nearest dwarf galaxies 
give detailed
information about how many stars formed, where, and when.
In general, star formation rates go up and down as time passes and the
galaxies evolve.
What is happening in the interstellar medium (ISM)
to cause these fluctuations in the star formation rate?
The issue is critical to an understanding of galaxy evolution.
We attempt to answer this question by studying the properties of the ISM
in a sample of dwarf galaxies with good color-magnitude information.

We also study the ISM in the nearby dwarfs because it may offer the 
closest
analog to conditions which prevailed in the very early universe.
Specifically, the nearby dwarfs have extremely low metallicities.
Heavy metals in the gas phase and in dust are responsible for 
most of the heating and cooling of the interstellar medium, not to mention
the formation of molecular gas. 
A better understanding of how the ISM works in the nearby dwarf galaxies
may find broader application to galaxy evolution and very distant systems.

\section{Sample and Observations}

It has long been known that a careful analysis of HI line profiles can
give insight into the temperature and density structure of the ISM
(e.g. Radhakrishnan et al. 1972).
We are applying these ideas to a sample of seven nearby ($\leq$ 3 Mpc)
gas-rich dwarf galaxies.
In order to study the HI line profiles without too much interference from
rotational broadening in the finite HI beam, the galaxies must be 
low mass and/or close to face-on.
The galaxies in our sample have HI masses about $10^7$ M$_\odot$, and
dynamical masses are a few times larger than that.
These galaxies have oxygen abundances in the range 4\% to 10\% of solar,
which are approaching the lowest metallicities known for galaxies.

The HI line profiles are obtained from VLA observations with 15\arcsec\ (70 to
200 pc) and 1.3 km/s resolution.
We also use ground-based optical broadband and H$\alpha$ imaging
(e.g. van Zee 2000) to show regions of current and past star formation activity.
Four of the galaxies have
stellar color-magnitude information from HST,
which allows us to trace the star formation activity as a function of time and place within
the galaxy.

\section{The Phase Structure of the Neutral ISM}


Contrary to expectations, the HI line width as measured by a second moment
of intensity {\it decreases} in regions of high column density and
near star formation activity.
This happens because the {\it shape}
of the HI line profile changes significantly near regions of star 
formation.
Throughout each galaxy studied, we find a broad HI component, well fitted by
a Gaussian of dispersion 8--10 km/s, and containing about 80\% of the HI in 
the galaxy.
Its dispersion limits its kinetic temperature to be $< 10^4$ K.
In some regions, near the locations of current star formation activity,
there is superposed a narrower component of dispersion 3--5 km/s, with about 20\%
of the HI mass and kinetic temperature $<$ 1000 K.
These temperature limits do not take into account possible line broadening
from magnetic fields, turbulence, and so on.
Figure 1 shows some of these types of spectra from the galaxy GR~8.
More details about the analysis of line profiles can be found
in Young \& Lo (1996; 1997), where we discuss the galaxies Leo~A, Sag~DIG, and LGS~3. 
 
\begin{figure}
\includegraphics[scale=0.65]{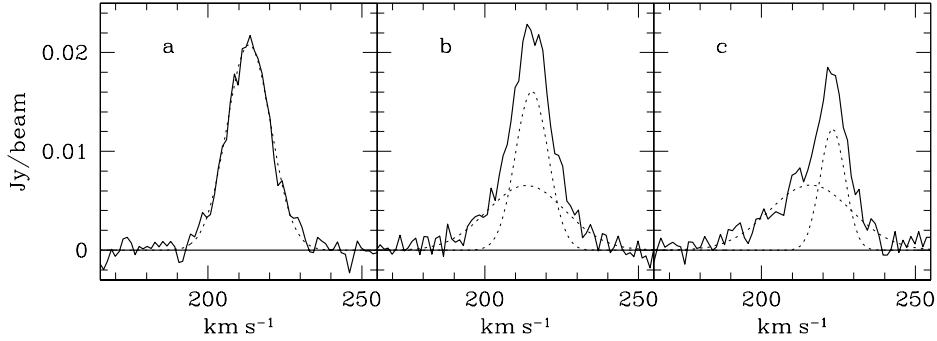}
\caption{HI spectra from the dwarf irregular galaxy GR 8.  
The solid line is the data and the dashed line shows the fitted Gaussian components.  
Panel a is a spectrum which is well described by the broad HI component alone.  
Panel b requires a broad and a narrow HI component; panel c shows two components
with velocities offset by about 5 km/s, possible evidence for kinetic energy
input into the ISM by young massive stars.
\label{spectrafig}}
\end{figure}

\begin{figure}
\includegraphics[keepaspectratio=true,scale=0.5,angle=-90]{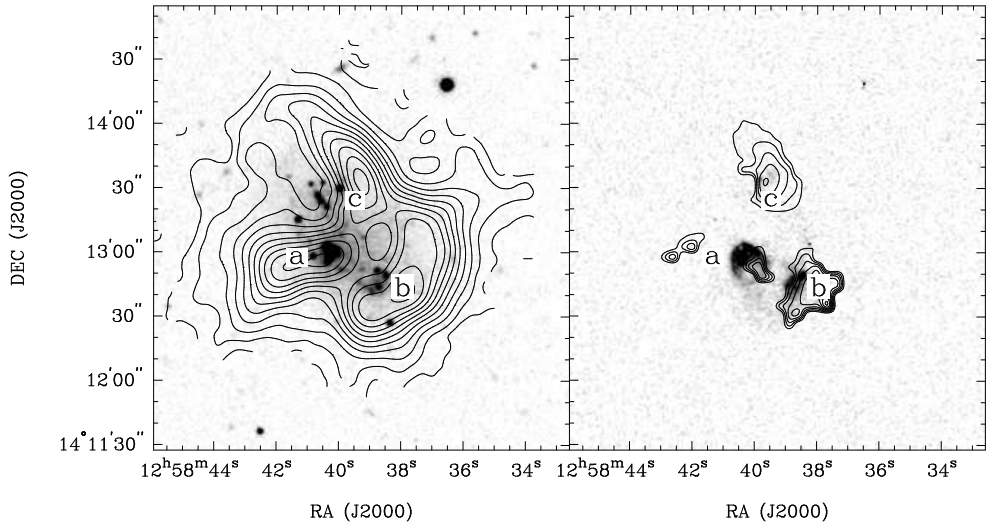}
\caption{GR 8: on the left, the contours show the total HI column density and the greyscale is
an optical broadband (R) image.  The letters a, b, and c indicate the
positions of the corresponding spectra in Figure \ref{spectrafig}.
The linear scale is 1\arcmin = 640 pc. 
On the right, the contours show the column density of just the narrow (cold)
HI component, and the greyscale is a continuum-subtracted H$\alpha$ image.
\label{opticalfig}}
\end{figure}


We identify the broader HI component as a warm (or more turbulent) neutral HI phase, and the
narrower component as a cold (or more quiescent) HI phase.
That's because the properties of the dwarf irregular HI spectra
are exactly analogous to those of the HI emission
spectra in our own Galaxy (Kulkarni \& Heiles 1988).
In addition,
the broad and narrow (warm and cold) HI features are also seen in
an intermediate velocity HI cloud (Shaw et al. 1996)
and the Magellanic Clouds (e.g. Dickey et al. 1994).
The broad and narrow HI components should be seen in many other
galaxies as well, if the sensitivity, spectral resolution, and no-rotational-broadening requirements can be met.

Because of the dependence of heating and cooling on heavy metals in the ISM,
as mentioned above,
it's not obvious whether a stable multiphase medium similar
to the one found in our Galaxy 
should exist in the dwarf irregulars.
Young \& Lo (1996, 1997) consider some theoretical models of the
phase structure of the Milky Way ISM.
They find that these Galactic models can be extended to conditions
which are appropriate for the dwarf irregulars, and the models are consistent
with observations if the thermal pressure in the dwarf irregulars is
a bit lower than in our own Galaxy.
These results give greater confidence that the Galactic ISM models
may be extended to fit other systems as well--- for example,
high-redshift quasar absorption lines also indicate multiphase media (Churchill, this conference; Chengalur, this conference).

\section{Star Formation and the ISM}

What do the warm and cold phases of the {\it atomic} medium have to do with
star formation?  
After all, the star formation is probably happening in {\it molecular} gas.
But CO emission has not been detected in galaxies of
this type (Taylor et al. 1998), presumably because of the low metallicity.  
We hypothesize that the
cold, dense atomic phase is telling us where conditions are right
for the formation of molecular gas and stars.
Figure \ref{opticalfig} shows that the narrow (cold) HI component is localized
near regions of star formation activity in GR~8, whereas the broad (warm) HI component
is spread throughout the galaxy.

The properties of the cold HI phase in the dwarf galaxies also explain
some important results about how star formation happens in these galaxies.
Dohm-Palmer et al. (1997, 1998) traced the star formation activity
in Sextans~A and in GR~8 as a function of both time and location.
They find that star formation occurs in regions or complexes of size 
100-200 pc; the activity lasts in one region for about 100 Myr 
and moves around through the galaxy in a way
which is loosely organized at best.
The narrow (cold) atomic component is 
found in patches of about the same size as the star formation regions
(Figure \ref{opticalfig}), so it may trace the raw material for star
formation activity. 
The sizes and linewidths of the narrow HI clumps imply virial masses
which are a few times larger than the HI masses alone;
part of the difference may be caused by molecular gas.

We plan to test these ideas more fully by exploring the evolution of the
ISM with time.
For example, the cold HI phase may be a prerequisite for star formation but 
should be gradually
destroyed as the young massive stars inject radiation and kinetic energy
into the surrounding gas.
We cannot watch the evolution of an individual region of a galaxy,
but stellar color-magnitude diagrams for a sample of galaxies will place 
ages on many different regions.
By finding a sequence of star formation regions of different ages, we
may be able to watch the evolution of the ISM through the onset of star 
formation and the disruption of the raw material for star formation.




\acknowledgements

J. Salzer is gratefully acknowledged for providing the optical and H$\alpha$
images of GR~8.

\vspace{2\baselineskip}
\noindent {\bf Question:} (J. Rhoads)
What are the dynamical times in these galaxies. ie., how long does it
take a star to orbit the galaxy and forget where it was born?

\vspace{\baselineskip}
\noindent {\bf Response:}
Dohm-Palmer et al. (1997) discuss this issue, and they argue that
the remnant structures of star formation activity could easily last
for 0.5 to 1 Gyr.  
One of the reasons for this longevity is that these small galaxies appear to be 
rotating in solid-body fashion, with no shear and no spiral density waves.

\end{document}